# Dynamic Range Reporting in External Memory

Yakov Nekrich*

Department of Computer Science, University of Bonn.

**Abstract**

In this paper we describe a dynamic external memory data structure that supports range reporting queries in three dimensions in $O(\log_B^2 N + \frac{k}{B})$ I/O operations, where $k$ is the number of points in the answer and $B$ is the block size. This is the first dynamic data structure that answers three-dimensional range reporting queries in $\log_B^{O(1)} N + O(\frac{k}{B})$ I/Os.

## 1 Introduction

The orthogonal range reporting problem is to maintain a set of points $S$ in a data structure so that for an arbitrary query rectangle $Q$ all points in $Q \cap S$ can be reported. This is a fundamental problem with several important applications, such as geographic information systems, computer graphics, and databases. In this paper we present a dynamic external-memory data structure that supports three-dimensional range reporting queries in $O(\log_B^2 N + \frac{k}{B})$ I/O operations and updates in $O(\log_2^3 N)$ I/O operations, where $k$ is the number of reported points and $N$ is the number of points in the data structure.

In the external memory model the data is stored in *disk blocks* of size $B$, a block can be read into internal memory from disk (resp. written from internal memory into disk) with one I/O operation, and computation can only be performed on data stored in the internal memory. The space usage is measured in the number of blocks, and the time complexity is measured in the number of I/O operations. A more detailed description of the external memory model can be found in e.g. [21] or [4]. Since we are interested in minimizing the number of I/O operations, an efficient data structure should support queries in $\log_B^{O(1)} N + O(\frac{k}{B})$ I/O operations.

In the RAM computation model, there are both static and dynamic data structures that use $N \log_2^{O(1)} N$ space and support $d$-dimensional orthogonal queries in $O(\log_2 N + k)$ time; see e.g., [3] for a survey of previous results. In the external memory model, these results can be matched only in two dimensions (dynamic data structure) and three dimensions (static data structure). The dynamic data structure of Arge *et al.* [9] uses $O((N/B) \log_2 N / \log_2 \log_B N)$ blocks of space and supports two-dimensional range reporting queries and updates in $O(\log_B N + \frac{k}{B})$ and $O(\log_B N (\log_2 N / \log_2 \log_B N))$ I/O operations respectively. The static data structure of Vengroff and Vitter [22, 21] supports three-dimensional range reporting queries in $O(\log_B N + \frac{k}{B})$ I/Os and uses $O((N/B) \log_2^4 N)$ blocks of space. The space usage of a three-dimensional data structures was improved by Afshani [1] and Afshani, Arge, and Larsen [2] to $O((N/B) \log_2^3 N)$ and $O((N/B)(\log_2 N / \log_2 \log_B N)^3)$ blocks respectively, see Table 1. The query cost can be improved if all point coordinates are positive integers bounded by a parameter $U$ [15, 16, 1], and the space usage can be reduced for some special cases of orthogonal queries, such as dominance queries; we refer the reader to [1, 2] for a more detailed description of special cases and to [7] for an extensive description of previous results.

---
*Email `yasha@cs.uni-bonn.de`.

| Model | Ref. | Query | Space | Update |
|---|---|---|---|---|
| $d=2$: | | | | |
| RAM | [13] | $O(\log_2 N/\log_2\log_2 N + k)$ | $O((N/B)\log^\omega N)$ | $O(\log^\omega N)$ |
| IO | [9] | $O(\log_B N + \frac{k}{B})$ | $O((N/B)\log_2 N/\log_2\log_B N)$ | $O(\log_B N \log_2 N/\log_2\log_B N)$ |
| $d=3$ | | | | |
| RAM | [13] | $O((\log_2 N/\log_2\log_2 N)^2 + k)$ | $O((N/B)\log^{\omega+1} N)$ | $O(\log^{\omega+1} N)$ |
| IO | [22] | $O(\log_B N + \frac{k}{B})$ | $O((N/B)\log_2^4 N)$ | - |
| IO | [1] | $O(\log_B N + \frac{k}{B})$ | $O((N/B)\log_2^3 N)$ | - |
| IO | [2] | $O(\log_B N + \frac{k}{B})$ | $O((N/B)(\log_2 N/\log_2\log_B N)^3)$ | - |
| IO | [2] | $O(\log_B N(\log_2 N/\log_2\log_B N) + \frac{k}{B})$ | $O((N/B)(\log_2 N/\log_2\log_B N)^2)$ | - |
| IO | [9]+[5] | $O(\log_B N(\log_2 N/\log_2\log_2 N) + \frac{k}{B})$ | $O((N/B)\log_2^{2+\varepsilon} N)$ | $O(\log_B N \log_2^{1+\varepsilon} N)$ |
| IO | * | $O(\log_B^2 N + \frac{k}{B})$ | $O((N/B)\log_2^2 N \log_2^2 B)$ | $O(\log_2^3 N)$ |

Table 1: Upper bounds for orthogonal range reporting in RAM and external memory models in two and three dimensions. Only dynamic results in the RAM model are listed. For comparison, the space usage of data structures in the RAM model is specified in blocks of size $B$. We denote by $\omega$ and $\varepsilon$ arbitrary constants such that $\varepsilon > 0$ and $\omega > 7/8$; our result is marked with an asterisk.

Using range trees with fan-out $\log^\varepsilon N$ [5], we can transform a two-dimensional data structure into a data structure that supports $d$-dimensional orthogonal queries, so that the cost of queries and updates increases by a $O(\log_2 N/\log_2\log_2 N)$ factor for each dimension and the space usage increases by a factor $O(\log_2^{1+\varepsilon} N)$ for each dimension. The recent (static) dimension reduction technique of [2] increases the cost of queries by $O(\log_2 N/\log_2\log_B N)$ factor and the space usage also by a $O(\log_2 N/\log_2\log_B N)$ factor. These techniques can be used to obtain three-dimensional data structures that support queries with $O(\log_B N(\log_2 N/\log_2\log_2 N) + \frac{k}{B})$ and $O(\log_B N(\log_2 N/\log_2\log_B N) + \frac{k}{B})$ I/Os respectively; see Table 1. However, these data structures do not achieve $O(\log_B^c N)$ query bound for any $B$ and a constant $c$. In the case when $B = \Omega((\log_2 N)^{f(N)})$ for some function $f(N) = \Omega(1)$, we need $\Omega(f(N)\log_B^2 N) + O(\frac{k}{B})$ operations to answer queries using the combination of [9] and [5] or the result of [2]. We also don't know if there are efficient (static or dynamic) data structures for range reporting in $d \geq 4$ dimensions that report all points with $\log_B^{O(1)} N + O(\frac{k}{B})$ operations.

In this paper we describe a data structure that uses $O(\frac{N}{B}\log_2^2 N \log_2^2 B)$ blocks of space, supports updates in $O(\log_2^3 N)$ amortized I/Os, and answers three-dimensional orthogonal range reporting queries in $O(\log_B^2 N + \frac{k}{B})$ I/Os. Thus our result "matches" the query complexity of the dynamic RAM data structure of [13]. Moreover, the space usage of our data structure differs by a $O(\log_2^2 B(\log_2\log_B N)^3/\log_2 N)$ factor from the best previously known external memory static data structure [2]. Hence, when $B$ is not very large, i.e., when $\log_2 B = o(\sqrt{\log_2 N/(\log_2\log_B N)^3})$, our dynamic data structure uses less space than the static data structure of [2].

In section 2 we describe the dynamic data structure that supports dominance queries in $O(\frac{k}{B})$ I/Os when the set $S$ contains $O(B^{4/3})$ points. Our data structure maintains $O(\log_2 B)$ $t$-approximate boundaries of [22], that will be defined in section 2. We show that each $t$-approximate boundary can be constructed with $O(B\log_2 B)$ I/O operations for $\geq B$ and a small set $S$. The cost of re-building the data structure is distributed among $O(B^{4/3})$ updates with the lazy updates approach: the newly inserted and deleted points are stored in two buffers for each $t$-approximate boundary, and each $t$-approximate boundary is re-built when one of its buffers contains the sufficient number of points. We further improve the update time by showing how to store only two buffers for all boundaries. The trick of storing inserted (deleted) points for different boundaries in the same buffer may be of independent interest. Using standard techniques, more general orthogonal range queries can be reduced to dominance queries as described in section 2.1.

In section 3 we describe the data structure that supports $(2,1,2)$-sided queries $Q = [a,b] \times [c,+\infty) \times [d,e]$ on a set of points $S$ such that $p.z = O(B^f)$ for a small constant $f$ and for any $p \in S$. Here and further we denote by $p.x$, $p.y$, and $p.z$ the $x$-, $y$-, and $z$-coordinates of a point $p$. The main idea of section 3 is to store points in a data structure $\mathcal{T}$ that is similar to the external memory priority search tree, but contains three-dimensional points. The data structure for small sets from section 2.1 is used to guide the search in each node of $\mathcal{T}$. The data structure that supports arbitrary $(2,1,2)$-sided queries is described in section 4. The data structure is based on a range tree with fan-out $\Theta(B^f)$ for a small constant $f$ that is built on $z$-coordinates of points. The main idea of section 4 is to store the data structure $F_v$ of section 3 in every node $v$ of the range tree. The $z$-coordinate of each point $p$ in $F_v$ is replaced with an index bounded by $\Theta(B^f)$ that indicates which child of the node $v$ contains $p$. We show how a general $(2,1,2)$-sided query can be reduced to $O(\log_B N)$ queries to data structures $F_v$. Finally, we can obtain a data structure for general three-dimensional queries from the data structure for $(2,1,2)$-sided queries using standard techniques.

Thus our approach is based on a combination of some previously known techniques with some novel ideas. In particular we believe that the data structures described in sections 3 and 4 and the general decomposition of the three-dimensional range reporting problem into subproblems are new.

## 2 Dominance Reporting for Small Sets

A point $q$ dominates a point $p$ if all coordinates of $q$ are greater than or equal to the respective coordinates of $p$. The dominance reporting query is to report all points $p \in S$ that dominate a query point $q$. A three-dimensional dominance reporting query is equivalent to reporting all points in a product of three half-open intervals. In this section we describe a dynamic data structure that contains $O(B^{4/3})$ elements and supports dominance reporting queries and updates. The main idea of this data structure is that the $t$-approximate boundary [22] for a small set of elements can be efficiently maintained under insertions and deletions.

**Overview.** A three-dimensional $t$-approximate boundary was introduced by Vengroff and Vitter [22]. A $t$-approximate boundary for a three-dimensional set $S$ is a surface $\mathcal{V}$ that satisfies the following properties: (1) $\mathcal{V}$ divides the space, i.e. every point either dominates a point on $\mathcal{V}$ or is dominated by a point of $\mathcal{V}$; (2) every point of $\mathcal{V}$ is dominated by at least $t$ and at most $3t$ points of $S$. An example of a $t$-approximate boundary constructed with the algorithm of [22] is shown on Fig. 1. There are $O(|S|)$ points on $\mathcal{V}$ called *inward corners*, such that every point on $\mathcal{V}$ dominates an inward corner and an inward corner does not dominate any point on $\mathcal{V}$ (except of itself). There is a linear space data structure that finds an inward corner $c$ of $\mathcal{V}$ that is dominated by a query point $q$, if such inward corner $c$ exists, and reports all points of $S$ that dominate $c$ in $O(\log_B(|S|) + t/B)$ I/Os. We maintain $(\log_2 B)/6$ $t$-approximate boundaries $\mathcal{V}_1, \mathcal{V}_2, \ldots, \mathcal{V}_s$, where $\mathcal{V}_i$ is a $B \cdot 2^{2i}$-approximate boundary. Given a query point $q$, we examine $\mathcal{V}_1, \mathcal{V}_2, \ldots, \mathcal{V}_i$ and find the minimal index $i$, such that $q$ dominates an inward corner $c_j$ of $\mathcal{V}_i$ using the method described in [22]. We can test each $\mathcal{V}_i$ in $O(\log_B B^{4/3}) = O(1)$ I/Os and find the index $i$ in $O(i)$ I/Os. If $q$ dominates an inward corner $c_j$ of $\mathcal{V}_i$ but does not dominate any point on $\mathcal{V}_{i-1}$, then $q$ is dominated by $\Theta(2^{2i}B)$ points of $S$. Since $q$ dominates $c_j$, all points that dominate $q$ also dominate $c_j$. Hence, we can examine the list of points that dominate $c_j$ and report all points that dominate $q$ in $O(2^{2i}) = O(\frac{k}{B})$ I/O operations. Thus the total query cost is $O(\frac{k}{B})$. See [22] for a more detailed description.

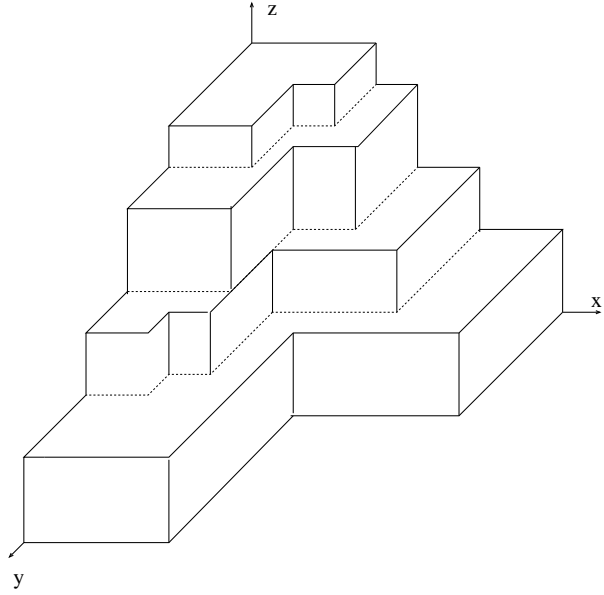

Figure 1: An example of a $t$-approximate boundary. The points of the set $S$ are not shown. Ridges $R'_2$, $R'_3$, $R'_4$, and $R'_5$ are drawn with dotted lines. Ridges $R_1$, $R_2$, $R_3$, and $R_4$ are drawn with solid lines. $A, B, C, D, E$ are examples of inward corners. $X, Y, Z$, and $W$ are examples of in-corners; $X$ belongs to ridge $R_1$, and $Y$, $Z$, and $W$ belong to ridge $R_3$.

We can construct a $t$-approximate boundary $\mathcal{V}_i$ with $O(B)$ I/O operations if $S$ contains $O(B^{4/3})$ points and $t \geq B$; the algorithm is described in section 5. In the next part of this section we show how the data structure for a small set of points can be dynamized by distributing the construction cost among $\Theta(B)$ update operations. This is achieved by storing buffers with newly inserted and deleted points and periodically rebuilding the data structure. Then, we show that we can support update operations in $O(1)$ I/Os on the data structure that consists of $O(\log_2 B)$ boundaries by storing one buffer with recently inserted points and one buffer with recently deleted points for *all* $t$-approximate boundaries.

**Deletion-only Data Structure.** A $t$-approximate boundary $\mathcal{V}_i$ supports lazy deletions in $O(1)$ amortized I/O operations. When a point $p$ is deleted, we simply add it to a list $\mathcal{D}$ of deleted elements that may contain up to $2^{2i-1}B$ points. Let $T$ be the list of points that dominate a query point $q$; we can obtain $T$ in $O(\frac{|T|}{B})$ I/Os as described in the beginning of this section. We can sort $T$ in $O(\frac{|T|}{B} \log_B |T|) = O(2^{2i} \log_B(2^{2i}B)) = O(2^{2i})$ I/Os (we assume that each point in $S$ has a unique integer identifier). We can also sort $\mathcal{D}$ in $O(2^{2i})$ I/Os. Then, we traverse $T$ and $\mathcal{D}$ and remove from $T$ all points that occur in $\mathcal{D}$ in $O(\frac{|T|+|\mathcal{D}|}{B}) = O(2^{2i})$ I/Os. Since we use $\mathcal{V}_i$ when $\frac{k}{B} = \Omega(2^{2i})$, the query cost remains unchanged. When the number of deleted points in $\mathcal{D}$ equals to $B \cdot 2^{2i}/2$, we re-build the data structure for $\mathcal{V}_i$ without deleted points in $O(B)$ I/Os and empty the list $\mathcal{D}$.

**Supporting Insertions.** Insertions can be supported with a similar technique. Inserted points are stored in the list of new points $\mathcal{I}$ that may contain up to $2^{2i-1}B$ points. When a point $p$ is deleted, we add it to a list $\mathcal{D}$ of deleted points as described above. If a point $p$ stored in $\mathcal{I}$ is deleted, we simply remove $p$ from $\mathcal{I}$. When $\mathcal{I}$ contains $2^{2i-1}B$ points, we re-build the data structure for $\mathcal{V}_i$. To answer a query, we examine all points from $T$ that do not belong to $\mathcal{D}$ in $O(\frac{k}{B})$ I/Os as described in the previous paragraph. Then, we traverse the list $\mathcal{I}$ and report all point that dominate the query point in $O(2^{2i-1}) = O(\frac{k}{B})$ I/Os.

**Updates with $O(1)$ Cost.** Since our data structure consists of $O(\log_2 B)$ boundaries $\mathcal{V}_i$, the total cost of an update is $O(\log_2 B)$. We can reduce the amortized update cost to a constant by storing new inserted points for all boundaries in one list $\mathcal{I}$ and deleted points for all boundaries in one list $\mathcal{D}$. An array $D$ stores pointers to elements of $\mathcal{D}$, such that all elements between $D[i]$ and the end of $\mathcal{D}$ are removed from the data structure for $\mathcal{V}_i$. An array $I$ stores pointers to elements of $\mathcal{I}$, such that all elements between $I[i]$ and the end of $\mathcal{I}$ are new elements that are not yet inserted into the data structure for $\mathcal{V}_i$. The pointer $\text{end}(\mathcal{D})$ ($\text{end}(\mathcal{I})$) points to the last (in chronological order) deleted (inserted) element stored in $\mathcal{D}$ ($\mathcal{I}$). Both $\mathcal{D}$ and $\mathcal{I}$ also contain one additional dummy element $l_D$ (resp., $l_I$) that follows $\text{end}(\mathcal{D})$ (resp., $\text{end}(\mathcal{I})$). When a new point $p$ is inserted, we store $p$ in the $l_I$, set the pointer $\text{end}(\mathcal{I})$ so that it points to $l_I$, and append a new dummy element after $\text{end}(\mathcal{I})$. A deleted element is appended at the end of $\mathcal{D}$ with the same procedure. After $2^{2i-1}B$ deletions we rebuild the data structure for $\mathcal{V}_i$ without deleted elements and change $D[i]$ so that it points to $l_D$. After $2^{2i-1}B$ insertions we rebuild the data structure for $\mathcal{V}_i$ with new elements and change $I[i]$ so that it points to $l_I$. After $\Theta(\log_2 B \cdot B)$ updates, we re-build the data structures for all $\mathcal{V}_i$ as well as the lists $\mathcal{I}$ and $\mathcal{D}$. This incurs an amortized cost $O(1)$. The total cost of re-building data structures and (pointers to) lists $\mathcal{D}$ and $\mathcal{I}$ in a sequence of $B^{4/3}$ update operations is $O(\sum_{j=0}^{r} 2^{r-j} B) = O(B^{4/3})$, where $r = \log_2 B/3 + \text{const}$ is the index of the last $t$-approximate boundary $\mathcal{V}_i$. We can report all points that dominate an inward corner of $\mathcal{V}_i$ in $O(2^{2i})$ I/Os as described above. Hence, dominance queries can be supported in $O(\frac{k}{B})$ I/Os. This result can be summarized in the following Lemma.

**Lemma 1** *Elements of a set $S$ such that $|S| = O(B^{4/3})$ can be stored in a data structure that uses $O(\frac{|S|}{B} \log_2 |S|)$ blocks of space and supports dominance queries in $O(\frac{k}{B})$ I/O operations and updates in $O(1)$ I/O operations amortized.*

## 2.1 $(1,1,2)$- and $(2,1,2)$-Sided Queries for Small Sets

Suppose that $b_x$, $b_y$, and $b_z$ are natural constants such that $1 \le b_x, b_y, b_z \le 2$. We say that a query $Q$ is a $(b_x, b_y, b_z)$-sided query if the projection of $Q$ on the $x$-axis is bounded on $b_x$ sides, the projection of $Q$ on the $y$-axis is bounded on $b_y$ sides and the projection of $Q$ on the $z$-axis is bounded on $b_z$ sides. Thus the projection of $Q$ on the $x$-axis (resp., $y$- or $z$-axis) is a an infinite half-open interval if $b_x$ (resp., $b_y$ or $b_z$) equals 1, and the projection of $Q$ on the $x$-axis (resp., $y$- or $z$-axis) is a finite closed interval if $b_x$ (resp., $b_y$ or $b_z$) equals 2. Dominance queries considered in section 2 are equivalent to $(1,1,1)$-sided queries. Using a standard reduction [11, 20], we can transform a $O(s(N))$ space data structure that supports $(1,1,1)$-sided queries in $O(t(N) + k/B)$ time and updates in $O(u(N))$ time into a $O(s(N) \log_2^m N)$ space data structure that supports $(b_x, b_y, b_z)$-sided queries in $O(t(N) + \frac{k}{B})$ time and updates in $O(u(N) \log_2^m N)$ time; here $m = b_x + b_y + b_z - 3$. Applying this transformation to Lemma 1, we obtain the following result.

**Lemma 2** *Let $1 \le b_x, b_y, b_z \le 2$ and $m = b_x + b_y + b_z - 3$. Elements of a set $S$ such that $|S| = O(B^{4/3})$ can be stored in a data structure that uses $O(\frac{|S|}{B} \log_2^{m+1} |S|)$ blocks of space and sup-*

ports $(b_x, b_y, b_z)$-sided queries in $O(\frac{k}{B})$ I/O operations and updates in $O(\log_2^m(|S|))$ I/O operations amortized.

In particular, we can support $(2,1,2)$-sided queries in $O(\frac{k}{B})$ I/Os and updates in $O(\log_2^2 B)$ I/Os on a set $S$ that contains $\Theta(B^{4/3})$ points using a data structure that needs $O(B^{1/3} \log_2^3 B)$ blocks of space.

## 3  Extended Three-Sided Queries

In this section we describe a data structure that supports $(2,1,2)$-sided reporting queries when $z$-coordinates of all points are positive integers bounded by $\Theta(B^f)$, $p.z = \Theta(B^f)$ for all points $p \in S$. Here $f$ is a constant such that $f \leq 1/6$

**Data Structure.** Our data structure is a modification of the external memory priority search tree [9]. The (external) priority search tree is a tree built on $x$-coordinates of two-dimensional points. A point stored in a leaf is associated with an ancestor of $l$ or with $l$ itself, so that the following property is guaranteed: points associated with a node $v$ have larger $y$-coordinates than points associated with descendants of $v$. The main idea of our modification is to maintain this property for every possible value of the $z$-coordinate. Then, we maintain the data structure of section 2.1 in each tree node and use it to guide the search, i.e., to decide which descendants of a node must be visited.

We construct a tree $\mathcal{T}$ with fan-out $\Theta(B^f)$ on the set of $x$-coordinates of all points. We store $\Theta(B^{1+f})$ values, i.e., $x$-coordinates of $\Theta(B^{1+f})$ consecutive points of $S$, in each leaf node. The range of an internal node $v$ is an interval $rng(v) = [a_v, b_v]$, where $a_v$ and $b_v$ are the smallest and the largest values stored in the leaf descendants of $v$.

We associate a set of points $S_v$ with each node $v$ of $\mathcal{T}$. Sets $S_v$ can be constructed by visiting nodes of $\mathcal{T}$ in pre-order. For the root $r$ of $\mathcal{T}$, let $L_r$ be the set of all points in $S$ sorted in increasing order by their $y$-coordinates, and let $L_r[j]$ be the set of all points $p \in S$, $p.z = j$, sorted in increasing order by their $y$-coordinates. The set $S_r[j]$ contains the last $B$ points of $L_r[j]$, i.e., $B$ points with largest $y$-coordinates. For each non-root node $v$ of $\mathcal{T}$, the list $L_v$ contains all points $p$ such that $p.x$ belongs to the range of $v$ and $p$ does not belong to any $S_w$, where $w$ is an ancestor of $v$; points in $L_v$ are sorted in increasing order by their $y$-coordinates. The list $L_v[j]$ contains all points $p \in L_v$ such that $p.z = j$. If $v$ is an internal node, the set $S_v[j]$ contains the last $B$ points of $L_v[j]$. If $v$ is a leaf, then $S_v[j]$ contains all points from $L_v[j]$. Note that $L_v[j]$ and $S_v[j]$ may contain less than $B$ points or even be empty for some $j$. The set $S_v$ is the union of all sets $S_v[j]$, $S_v = \cup_j S_v[j]$. For any node $v$, $|S_v| = O(B^{1+f})$ The set $S'_v$ contains at most one point from each set $S_v[j]$. If $|S_v[j]| = B$, then $S'_v$ contains the point $p \in S_v[j]$ with minimal $y$-coordinate; otherwise $S'_v$ contains no points from $S_v[j]$.

We store data structures $D_v$ and $D'_v$ in each internal node $v$ of $\mathcal{T}$. The data structure $D_v$ contains all points of $S_{v_i}$ for every child $v_i$ of $v$, and the data structure $D'_v$ contains all points of $S'_{v_i}$ for every child $v_i$ of $v$. Thus $D_v$ contains $O(B^{1+2f})$ points, and $D'_v$ contains $O(B)$ points. By Lemma 2, $D_v$ and $D'_v$ can be stored in $O(B^{2f} \log_2^3 B)$ and $O(\log_2^3 B)$ blocks respectively and support $(2,1,2)$-sided queries in $O(1)$ I/O operations. In every node $v$ of $\mathcal{T}$, we also store a data structure $E_v$ that contains all points of $S_v$ and supports $(2,1,2)$-sided queries. Note that lists $L_v$ and $L_v[j]$ and sets $S_v[j]$ are not stored in the data structure; we only use them to simplify the description.

**Search Procedure.** Given a query $Q = [a,b] \times [c, +\infty) \times [d,e]$, we identify leaves $l_a$ and $l_b$: $l_a$ contains the smallest value that is greater than $a$ and $l_b$ contains the largest value that is smaller than $b$. Let $\pi_a$ and $\pi_b$ denote the paths from the root of $\mathcal{T}$ to $l_a$ and $l_b$ respectively. Let $\pi = \pi_a \cup \pi_b$ denote the set of all nodes of $\mathcal{T}$ that belong to $\pi_a$ or $\pi_b$. Every point $p \in S$ such that $p.x \in [a,b]$ is stored in some set $S_v$ such that either $v$ belongs to $\pi$ or $v$ is a descendant of a node that belongs to $\pi$.

We can visit all nodes $v \in \pi$ and report all points in $S_v \cap Q$ in $O(\log_B N)$ I/Os using data structures $E_v$ (we ignore the time needed to output points). Points in descendants of $v \in \pi$ can be found using the following Property.

**Fact 1** *Let $v'$, $v' \notin \pi$, be a child of a node $v \in \pi$, and let $w$ be a descendant of $v'$. If $S_w[j] \cap Q \neq \emptyset$, then $|S_{\mathrm{par}(w)}[j] \cap Q| = B$ where $\mathrm{par}(w)$ denotes the parent of a node $w$.*

*Proof*: Recall that $Q = [a,b] \times [c,+\infty) \times [d,e]$. For a child $v'$ of $v$, such that $v' \notin \pi$, either $rng(v') \cap [a,b] = \emptyset$ or $rng(v') \subset [a,b]$. Hence, Fact 1 is non-trivial only in the case when $j \in [d,e]$ and $rng(v') \subset [a,b]$. In this case a point $p \in S_w[j]$ (resp., $p \in S_{\mathrm{par}(w)}[j]$) belongs to $Q$ if and only if $p.y \geq c$. Suppose that some $p \in S_w[j]$ belongs to $Q$. Since $p.y \geq c$ and $p'.y > p.y$ for any point $p' \in S_{\mathrm{par}(w)}[j]$, all points $p' \in S_{\mathrm{par}(w)}[j]$ belong to $Q$. The set $S_{\mathrm{par}(w)}[j]$ contains $B$ points because $S_w[j]$ is not empty. $\square$

Consider a node $v$, such that $v \in \pi_a$ and $v \notin \pi_b$. Suppose that the $i$-th child $v_i$ of $v$ belongs to $\pi_a$ and $rng(v_{i+1}) = [a',b']$. We define the query $Q_v = [a',b] \times [c,+\infty) \times [d,e]$. For any point $p$ stored in a descendant $w$ of $v$, such that $w \notin \pi_a$, queries $Q_v$ and $Q$ are equivalent: $p$ belongs to $Q$ if and only if $p$ belongs to $Q_v$. Points in $S_w \cap Q = S_w \cap Q_v$ for all descendants $w$ of $v$, $w \notin \pi_a$, can be reported with the following recursive procedure. We report all points in $Q_v \cap S_{v_i}$ for all children $v_i$ of $v$ using the data structure $D_v$. All children $v_i$ of $v$, such that $Q_v \cap S_{v_i}[j]$ contains at least $B$ points for at least one $j$, can be identified using $D'_v$. We visit all such non-leaf nodes $v_i$ and recursively call the same procedure.

Our procedure reports all points in $S_w \cap Q_v$: Suppose that $S_w[j] \cap Q_v \neq \emptyset$ for some $w$ and $j$. Then $S_{\mathrm{par}(w)}[j] \cap Q_v$ contains $B$ points by Fact 1. Hence, the parent of $w$ will be visited and all points in $S_w \cap Q_v$ will be reported by querying the data structure $D_{\mathrm{par}(w)}$. If $k_v$ is the total number of points in $S_w \cap Q_v$ for all $w$, then the search procedure takes $O(\frac{k_v}{B})$ I/O operations: Queries answered by $D_w$ and $D'_w$ in every visited node $w$ take $O(1)$ I/O operations and a node $w$ is visited only if $|S_w[j] \cap Q_v| = B$ for at least one value of $j$.

All points in $S_w \cap Q$ for all descendants $w$ of a node $v$, such that $v \in \pi_b$ but $v \notin \pi_a$ or $v$ is the lowest common ancestor of $l_a$ and $l_b$, can be found with the same procedure. The only difference is that the query $Q_v$ is defined differently: if $v \in \pi_b$, $v \notin \pi_a$, and the $i$-th child $v_i$ of $v$ belongs to $\pi_b$, then $Q_v = [a,b'] \times [c,+\infty) \times [d,e]$ where $rng(v_{i-1}) = [a',b']$. If $v$ is the lowest common ancestor of $l_a$ and $l_b$, then $v \in \pi_a$ and $v \in \pi_b$. Suppose that $v_i \in \pi_a$ and $v_l \in \pi_b$ where $v_i$ and $v_l$ are the children of $v$. Then $Q_v = [a',b''] \times [c,+\infty) \times [d,e]$ where $rng(v_{i+1}) = [a',b']$ and $rng(v_{l-1}) = [a'',b'']$. Hence, a query $Q$ can be answered with $O(\log_B N + \frac{k}{B})$ I/O operations.

**Space Usage and Updates.** Every data structure $D_v$ contains $O(B^{1+2f})$ points and can be stored in $O(B^{2f} \log_2^3 B)$ blocks of space. Every $D'_v$ contains $O(B^{2f})$ points and can be stored in $O(\log_2^3 B)$ blocks. There are $O(\frac{N}{B^{1+2f}})$ internal nodes in $\mathcal{T}$; hence, all $D_v$ and $D'_v$ use $O(\frac{N}{B} \log_2^3 B)$ blocks. Every data structure $E_v$ contains $O(B^{1+f})$ points. Since the total number of nodes is $O(\frac{N}{B^{1+f}})$, all $E_v$ can be stored in $O(\frac{N}{B} \log_2^3 B)$ blocks.

When a point $p$ is inserted into $S$, we identify the leaf $l_p$ in which $p.x$ must be stored and traverse the path $\pi_p$ from $l_p$ to the root until we find a node $v$ such that $p.y < m_v.y$ and $m_v$ is the point with maximal $y$-coordinate in $S_v[p.z]$. Then, we insert $p$ into $S_v[p.z]$. Now $S_v[p.z]$ may contain $B+1$ points; if $|S_v[p.z]| = B+1$, the point $s_v$ with the smallest $y$-coordinate must be removed from $S_v[p.z]$. We insert the point $s_v$ into $S_{v_i}[p.z]$ where $v_i$ is the child of $v$ such that $v_i$ belongs to $\pi_p$. If $S_{v_i}[p.z]$ contains $B+1$ points, we move the point with the smallest $y$-coordinate from $S_{v_i}[p.z]$ to $S_u[p.z]$ where $u$ is the child of $v_i$, $u \in \pi_p$. The procedure continues until $S_u[p.z]$ contains at most $B$ points or the leaf $l_p$ is reached. In every node $u$ visited by the insertion procedure, one point is inserted into $S_u$ and at most one point is deleted from $S_u$. Hence data structures $E_u$, $D_w$, and $D'_w$, where $w$ denotes the parent of $u$, can be updated in $O(\log_2^2 B)$ I/Os. Since $O(\log_B N)$ nodes are visited, insertion takes $O(\log_2 N \log_2 B)$ I/O operations. Deletions can be supported with a similar procedure.

It remains to show how the tree $\mathcal{T}$ can be re-balanced after update operations, so that the height of $\mathcal{T}$ is $O(\log_B N)$. We implement the base tree $\mathcal{T}$ as a WBB-tree [10] with leaf parameter $n_l = B^{1+1/f}$ and branching parameter $n_b = B^{1/f}$. In a WBB tree with this choice of parameters the following invariants are maintained: each leaf contains between $B^{1+1/f}$ and $2B^{1+1/f} - 1$ values and for each internal node $v$ on level $h$ (counting from the lowest level) there are between $B^{1+(h+1)/f}/2$ and $2B^{1+(h+1)/f} - 1$ values stored in leaf descendants of $v$. It is also shown in [10] that internal node has between $B^{1/f}/4$ and $4B^{1/f}$ children. Hence, the height of $\mathcal{T}$ is $O(\log_B N)$.

If the invariants of a WBB tree are violated after an insertion, i.e., if a node $v$ on level $h$ contains $2B^{1+(h+1)/f}$ values (resp., $v$ contains $2B^{1+1/f}$ values if $v$ is a leaf), then we split the node $v$ into $v'$ and $v''$ that contain $B^{1+(h+1)/f}$ ($B^{1+1/f}$) values each. Splitting a node does not affect the children of $v$, i.e., every child of $v$ becomes the child of $v'$ or $v''$ after splitting. It can be shown [10] that a node $v$ on level $h$ is split at most once when a sequence of $B^{1+(h+1)/f}/2$ values is inserted into leaf descendants of $v$. See [10] for a complete description of the splitting procedure.

When a node $v$ is split into $v'$ and $v''$, data structures in nodes $v'$, $v''$, and in their descendants may change. Since $S_v[j] = S_{v'} \cup S_{v''}$ for each $j$ after the split operation, at least one of $S_{v'}[j]$ and $S_{v''}[j]$ contains less than $B$ points. Suppose that for some $j$, the set $S_{v'}[j]$ contains less than $B$ elements. If $S_{v_i}[j] \neq \emptyset$ for at least one child $v_i$ of $v'$, then some points must be moved from sets $S_{v_t}[j]$ into $S_v[j]$, where $v_t$ is a child of $v'$. Let $d_v = \min(|\cup S_{v_t}[j]|, B - |S_v|)$. We can identify $d_v$ points with largest $y$-coordinates in $\cup S_{v_t}$, using $D_{v'}$ and insert those points into $S_{v'}[j]$. Data structures $E_v$, $D_w$, and $D'_w$ where $w$ is a parent of $v'$ are updated accordingly. If $d_v > 0$, we recursively check sets $S_{v_t}$ for all children $v_t$ of $v'$. Data structures stored in the node $v''$ and the descendants of $v''$ are processed in the same way. Each point is moved only once and the total number of moved points does not exceed the total number of values stored in leaf descendants of $v'$ and $v''$. When a point is moved, all affected data structures can be updated in $O(\log_2^2 B)$ I/Os. The number of values stored in a node $v$ on level $h$ and all its descendants is $\Theta(B^{1+(h+1)/f})$. Since $v$ is split at most once after $B^{1+(h+1)/f}/2$ operations, the amortized cost for splitting a node is $O(\log_2^2 B)$. Every leaf has $O(\log_B N)$ ancestors; hence, the total amortized costs of splits incurred by an inserted point is $O(\log_2 B \log_2 N)$. Thus the total cost of an insertion is $O(\log_2 N \log_2 B)$.

We implement deletions with the lazy deletions approach. Suppose that a point $p$ such that $p.x$ is stored in a leaf $l_p$ is deleted from $S$. Then we mark the value $p.x$ as deleted in $l_p$. When $N/2$ values stored in leaves of $\mathcal{T}$ are marked as deleted, we rebuild the tree $\mathcal{T}$ and all secondary data structures. This can be done in $O(N \log_2^2 B)$ I/O operations. Hence, rebuilding after deletions incurs an amortized cost of $O(\log_2^2 B)$.

The result of this section is summed up in the following Lemma.

**Lemma 3** *There exists a $O(\frac{N}{B} \log_2^3 B)$ space data structure that supports extended three-sided*

queries in $O(\log_B N + \frac{k}{B})$ I/O operations and updates in $O(\log_2 N \log_2 B)$ I/O operations.

## 4 Range Reporting in Three Dimensions

Using range trees with fan-out $\Theta(B^f)$, we can transform the result of section 3 into a data structure for $(2,1,2)$-sided queries. For completeness, we sketch the data structure below.

We construct an external memory range tree on $z$-coordinates of the points in a set $S$: $z$-coordinates of all points are stored in leaves of the tree; each leaf contains $\Theta(B)$ values and each internal node has $\Theta(B^f)$ children. We denote by $R_v$ the set of points whose $z$-coordinates are stored in leaf descendants of the node $v$. The data structure $F_v$ contains one point for each point $p \in R_v$. If $p = (p.x, p.y, p.z)$, $p \in R_v$, is also stored in the $i$-th child $v_i$ of $v$, then $F_v$ contains the point $p' = (p.x, p.y, i)$. In other words, we replace the $z$-coordinate of each point $p \in R_v$ with an index $i \in [1, \Theta(B^f)]$, such that $p \in R_{v_i}$. $F_v$ supports $(2,1,2)$-sided queries as described in Lemma 3.

For each internal node $v$, let $int(v, i, j)$ denote the interval $[\min_i, \max_j]$ where $\min_i$ denotes the minimal value stored in a leaf descendant of the $i$-th child of $v$, and $\max_j$ denotes the maximal value stored in a leaf descendant of the $j$-th child of $v$. For a query $Q = [a,b] \times [c, +\infty) \times [d, e]$, we can represent the interval $[d, e]$ as a union of $O(\log_B N)$ intervals $int(v, g_i, g_j)$. Hence, $Q$ can be answered by answering $O(\log_B N)$ queries of the form $[a,b] \times [c, +\infty) \times int(v, g_i, g_j)$. Every such query can be answered by the data structure $F_v$. Hence, a $(2,1,2)$-sided query can be answered with $O(\log_B^2 N + \frac{k}{B})$ I/O operations. Since each point is stored in $O(\log_B N)$ data structures $F_v$, the space usage and update cost increase by a factor $O(\log_B N)$ compared with the data structure of Lemma 3.

**Lemma 4** *There exists a $O(\frac{N}{B} \log_2 N \log_2^2 B)$ space data structure that supports $(2,1,2)$-sided queries in $O(\log_B^2 N + \frac{k}{B})$ I/Os and updates in $O(\log_2^2 N)$ I/Os amortized.*

Finally, we apply the reduction described in section 2.1 and obtain the main result of this paper. The space usage and update cost increase by a factor $O(\log_2 N)$ in comparison with Lemma 4

**Theorem 1** *There exists a $O(\frac{N}{B} \log_2^2 N \log_2^2 B)$ space data structure that supports three-dimensional orthogonal range reporting queries in $O(\log_B^2 N + \frac{k}{B})$ I/O operations and updates in $O(\log_2^3 N)$ amortized I/O operations.*

## 5 Construction of a $t$-Approximate Boundary.

We describe below a (slightly simplified) variant of the construction algorithm from [22] for the case when all points have different $x$-, $y$-, and $z$-coordinates. The algorithm constructs a series of ridges in order of descending $z$-coordinates. The ridge $R_0$ consists of a single point $(0, 0, z_{\max})$, where $z_{\max}$ is the maximum $z$-coordinate of a point in $S$. During the $i$-th iteration, $i = 1, \ldots$, the ridge $R_i$ is constructed as follows. We move down $R_{i-1}$ until some point on $R_{i-1}$ is dominated by $3t$ points or $R_{i-1}$ hits the $(x, y)$-plane; the new position of $R_{i-1}$ is denoted by $R'_i$. Let $p$ be the point of $R'_i$ that lies on the $(x, z)$-plane. We move $p$ in the $+x$ direction until $p$ is dominated by $2t$ points of $S$. Then, the following loop is repeated until $p$ hits the $(y, z)$-plane: (1) the $y$-coordinate of $p$ is increased until $p$ is dominated by $t$ points (2) the $x$-coordinate of $p$ is decreased until $p$ is dominated by $2t$ points or $p$ hits $R'_i$ (3) if $p$ hits the ridge $R'_i$, $p$ follows $R'_i$ until it hits the $(y, z)$ plane or $p$ is dominated by $2t$ points. The ridge $R_i$ is constructed when $p$ hits the $(y, z)$-plane.

Positions of $p$ before the loop begins and at the end of step (2) are called inner corners of $\mathcal{V}$. Points on $R'_i$ with the same $(x,y)$-coordinates as some inner corner on $R_{i-1}$ are also called inner corners. If $p$ is an inner corner of some $R'_i$ but $p$ is not an inner corner of $R_i$, then $p$ is an inward corner. As described above, if all inward corners of a $t$-approximate boundary $\mathcal{V}$ are known, then we can determine whether a query point $q$ dominates some point of $\mathcal{V}$.

A $t$-approximate boundary for a set $S$ consists of $O(\frac{|S|}{t})$ ridges: since some point of each $R'_i$ except of the lowest one is dominated by $3t$ points of $S$ and each point of $R_{i-1}$ is dominated by $2t$ points, there are at least $t$ points with $z$-coordinates between $R_{i-1}$ and $R'_i$. The number of inner corners on each ridge is also $O(\frac{|S|}{t})$: suppose that during step (2) point $p$ moves from position $q$ to position $r$, i.e. $q$ is reached during the previous step (1) and $r$ is the inner corner. Then there are $t$ points whose $x$-coordinates are between $r.x$ and $q.x$. If $t \geq B$ and $|S| \leq B^{4/3}$, then the number of ridges in a $t$-approximate boundary and the number of inner corners in each ridge is $O(B^{1/3})$. We can use this to construct the data structure for a $t$-approximate boundary with $O(B)$ operations.

**Lemma 5** *If $|S| = O(B^{4/3})$ and $t \geq B$, then a $t$-approximate boundary for $S$ can be constructed with $O(B^{2/3})$ I/O operations.*

*Proof*: All points of $S$ are sorted in decreasing order by their $z$-coordinates and stored in a list $L$. Suppose that the ridge $R_i$ is already constructed. We store the inner corners of $R_i$ in the data structure $\mathcal{R}$. The number of elements in $\mathcal{R}$ is $O(B^{1/3})$; hence, $\mathcal{R}$ can be stored in the main memory. For every element $e$ of $\mathcal{R}$ we store the number of already processed points in $L$ whose projections on the $(x, y)$-plane dominate the projection of $e$ on the $(x, y)$-plane. Processed points have higher $z$-coordinates than the current position of $R_i$. We read the next $B$ points from $L$ and look for the highest point $p$, such that some $e \in R$ is dominated by $3t$ points $q \in L$ with $q.z \geq p.z$. If there is no such $p$, we modify the data structure $\mathcal{R}$, decrease the $z$-coordinate of $R_i$, and read the next $B$ points from $L$. This step is repeated until we find a point $p$ dominated by $3t$ points. When $p$ is found, the $z$-coordinate of $R_i$ is set to $p.z$. Then we set $R'_{i+1} = R_i$ and proceed with construction of $R_{i+1}$. Every time when we read a block of $B$ points, we either process $B$ points in $L$ or construct a new ridge. Since the number of ridges is $O(|S|/t) = O(B^{1/3})$ and the number of point in $L$ is $O(B^{4/3})$, the total number of I/Os needed to process $L$ is $O(B^{1/3})$.

When the $z$-coordinate of a ridge $R_i$ is known, $R_i$ can be constructed in $O(B^{1/3})$ I/Os. We divide the already processed points of $L$ into groups of $B$ points sorted in decreasing order by their $x$-coordinates: all points in a group $G_j$ have larger $x$-coordinates than points in $G_{j+1}$. We can obtain all groups $G_i$ in $O(B^{1/3})$ I/O operations. We also divide the already processed points of $L$ into groups $Y_i$ of $B$ points sorted by their $y$-coordinates: all points in a group $Y_i$ have smaller $y$-coordinates than points in $Y_{i+1}$. Let $p$ be the point on $R'_i$ that lies on the $(x, z)$ plane (i.e., $p.y = 0$). We move $p$ in $+x$ direction until $p$ is dominated by $2t$ points and identify the starting point of $R_i$; this can be done in $O(B^{1/3})$ I/Os. Suppose that $x$-coordinates of all points in $G_1, G_2, \ldots, G_{j-1}$ are greater than $p.x$. We initialize the variable $h$ to $j$ and the variable $v$ to 1; we read $G_h$, $Y_v$, and the inner corners of $R'_i$ into the main memory. Observe that since a ridge has $O(B^{1/3})$ inner corners, all inner corners of $R_i$ and $R'_i$ can be stored in the main memory. We perform the steps (1)-(3) of the loop as long as the $x$-coordinate of $p$ is greater than or equals to the minimum $x$-coordinate of a point in $G_h$ and the $y$-coordinate of $p$ is smaller than or equals to the maximum $y$-coordinate of a point in $Y_v$. As long as those conditions are satisfied we can determine the number of points that dominate $p$ using $G_h$ and $Y_v$: when a point $p$ is moved in $+y$ direction, the number of points that dominate $p$ can be changed only because of points in $Y_v$; when a point $p$ is moved in $-x$ direction, the number of points that dominate $p$ can be changed only because of points in $G_h$. If $p.y$ is greater than the maximal coordinate in $Y_v$, we read the next $Y_{v+1}$ into main memory and increment $v$ by 1. If $p.x$ is smaller than the minimal coordinate of $G_h$, then, we read $G_{h+1}$ and increment $h$ by 1. Since there are $O(B^{1/3})$ groups $G_h$ and $Y_v$, the total number of I/O operations needed to construct a ridge is $O(B^{1/3})$. Since there are $O(B^{1/3})$ ridges, we need $O(B^{2/3})$ operations to construct all ridges. Hence, the total construction cost is $O(B^{2/3})$. □

Since there are $O(|S|)$ inward corners, we cannot directly store $\Theta(t)$ points that dominate each inward corner. A data structure that uses linear space and reports all points that dominate an arbitrary inward corner is described in [14]. We can transform the data structure of [14] into an external memory data structure $\mathcal{E}$; in the case when $|S| = O(B^{4/3})$ the data structure $\mathcal{E}$ uses $O(B^{1/3})$ blocks of space and reports all points that dominate an arbitrary inward corner in $O(\frac{t}{B})$ I/Os. The following lemma shows how we can support batches of range reporting queries on a small set.

**Lemma 6** *For any $c \geq 3$, there exists a data structure for a set of $F = O(B^{1+1/c})$ points that supports $f = F/B$ range reporting queries in $O(B^{1/c} + X/B)$ I/O operations where $X = \sum_{i=1}^{f} X_i + f$ and $X_i$ is the number of points in the answer to the $i$-th query. The data structure uses $B^{1/c}$ blocks of space and can be constructed in $O(B^{1/c})$ I/O operations.*

*Proof*: Suppose that a set $A$ consists of $F$ points. We divide $A$ into $F/B = O(B^{1/c})$ subsets $A_i$, such that each $A_i$ contain $B$ points. We can read all queries $Q_j$, $j = 1, \ldots, F/B$, and all points of $A_i$ into the main memory with $O(1)$ read operations. Then, we can find all pairs $(p, j)$, such that $p \in A_i \cap Q_j$. For all sets $A_i$, we need $O(B^{1/c} + \frac{X}{B})$ read and write operations to produce a list $L$ that contains all such pairs. The list $L$ contains all points that belong to $Q_1, \ldots, Q_{F/B}$. It remains to determine which points belong to which query ranges. Using e.g., the sorting algorithm of [8], we sort $L$ by its second component in $O(\frac{X}{B} \log_B X) = O(\frac{X}{B})$ operations. Now we can scan the list and output all points $p$ that belong to a pair $(p, i)$ as the answer to a query $Q_i$ for $i = 1, \ldots, f$. □

The data structure $\mathcal{E}$ can be constructed by constructing $B^{1/3}$ instances of Lemma 6 data structures and answering batches of range reporting queries [14]. We use three data structures $V^{\min}$, $V^{\max}$, and $X$ that support three-dimensional queries. All data structures answer $B^{1/3}$ batches of queries (one batch for each ridge), and each batch consists of $B^{1/3}$ queries. Each batch can be processed in $O(B^{1/3})$ I/O operations by Lemma 6. Hence the data structure $\mathcal{E}$ can be constructed in $O(B)$ I/Os. We consider the case when the set of points $S$ contains $O(B^{4/3})$ points, and $t \geq B$. The algorithm presented below and its description are very similar to the algorithm that will be included into the full version of [14]. The description in this section is provided only for the sake of completeness.

The inward corners $c_j$ of $\mathcal{V}$, such that $\pi(c_j)$ dominates $\pi(c_i)$ for some inward corner $c_i$, but does not dominate the projection on the $(x, y)$-plane of one of the previously visited inward corners are called *the children* of $c_i$. A corner $c_i$ is a parent of $c_j$ if $c_j$ is a child of $c_i$. Descendants of $c_i$ are children of $c_i$ and descendants of children of $c_i$. With each inward corner $c_i$ we associate a list of points $Dom(c_i)$. A $d$-neighbor of an inward corner $v$ is one of $d$ preceding or $d$ following corners on the same ridge. The dominance list $Dom(c_i)$ contains all points $p$ that satisfy the following conditions:

- $p$ dominates $c_i$

- $p$ is not contained in the dominance lists of descendants of $c_i$

- $p$ is not contained in the dominance lists of descendants of one of the 3-neighbors of $c_i$

- $p$ is not contained in the dominance lists of those 3-neighbors of $c_i$ that belong to ridges with lower $z$-coordinate

When dominance lists for all inward corners are constructed, we can output all points that dominate any inward corner of a $t$-approximate boundary in $O(t/B)$ operations. Details can be found in [14]. We will show below how dominance lists can be constructed in external memory with help of 6.

We construct dominance lists for inward corners of ridges $R'_g, \ldots, R'_1$, so that ridges are processed in the ascending order of their $z$-coordinates. For each ridge $R'_i$ inward corners are processed in the ascending order of their $x$-coordinates. We will use three auxiliary data structures: The static data structure $X$ stores all points of $S$ and supports three-dimensional range reporting queries. Data structures $V^{\max}$ and $V^{\min}$ support updates and three-dimensional dominance reporting queries. At the beginning of the algorithm data structures $V^{\max}$ and $V^{\min}$ are empty; we will update these data structures every time when all inward corners of some ridge $R'_i$ are processed.

Points in the dominance list of an inward corner $c_j \in R'_i$ can be divided into two groups: **1.** Points that are stored in a dominance list of some inward corner(s) $c_s$, such that $c_s$ is neither a 3-neighbor of $c_j$, nor a descendant of $c_j$ or one of its 3-neighbors. **2.** "New" points, i.e. points that dominate $c_j$ but do not dominate any previously processed inward corner $c_s$. An example is shown on Fig. 2.

Points in the first group can be found with help of data structures $V^{\max}$ and $V^{\min}$: when the dominance list of an inward corner $c_j \in R'_i$ is constructed, $V^{\max}$ and $V^{\min}$ contain information about all points stored in the dominance lists of inward corners on ridges $R'_g, R'_{g+1}, \ldots, R'_{i+1}$. For every point $p = (x_p, y_p, z_p)$, data structure $V^{\min}$ contains an element $(x_p, ind_l, z_p)$. Here $ind_l$ denotes the index of the inward corner $c_l$, such that $p$ belongs to $Dom(c_l)$. If $p$ is stored in the dominance lists of more than one inward corner, then we choose the corner $c_l$ with the highest index. Data structure $V^{\min}$ supports queries $(a^+, b^-, c^+)$: report all elements $(x_p, ind_p, z_p)$ of $V^{\min}$ such that $x_p \geq a^+$, $ind_p < b^-$, and $z_p \geq c^+$. Clearly, such queries are equivalent to three-dimensional dominance reporting queries. The data structure $V^{\max}$ contains a point $(x_p, ind_l, z_p)$ for every point $p = (x_p, y_p, z_p)$. Again $ind_l$ is the index of an inward corner $c_l$ with $p \in Dom(c_l)$, but if $p$ is stored in more than one dominance list, we choose the inward corner $c_l$ with the lowest index. $V^{\max}$ supports queries $(a^+, b^+, c^+)$: report all elements $(x_p, ind_p, z_p)$ of $V^{\max}$ such that $x_p > a^+$, $ind_p > b^+$, and $z_p > c^+$.

If a point $p$ belongs to the first group, then it is stored either in a dominance list of a (descendant of) $d$-predecessor of $c_j$ or in a dominance list of a (descendant of) $d$-successor of $c_j$ for $d \geq 4$. Let $\min_j$ be the minimal index of a descendant of a 3-predecessor of $c_j$. Let $\max_j$ be the maximal index of a descendant of a 3-successor of $c_j$. Let $c_j = (x_j, y_j, z_j)$ For each point $p$ that dominates $c_j$ and is stored in a dominance list of a $d$-predecessor of $c_j$ for $d > 3$, data structure $V^{\min}$ contains a point $(x_p, ind_l, z_p)$, such that $ind_l < \min_j$, $x_p \geq x_j$, and $z_p \geq z_j$. If for an element $(x_p, ind_l, z_p)$ of $V^{\min}$, $ind_l < \min_j$ and $x_p > x_j$, then $y_p > y_j$ for the corresponding point $p$: $p$ dominates an inward corner that dominates some $d$-predecessor $c_p$ of $c_j$, and the $y$-coordinate of $c_p$ is greater than the $y$-coordinate of $c_j$. Hence, every element $(x_p, ind_l, z_p)$ of $V^{\min}$, such that $ind_l < \min_j$, $x_p \geq x_j$, and $z_p \geq z_j$ corresponds to a point that dominates $c_j$. All such elements can be found by a query to $V^{\min}$.

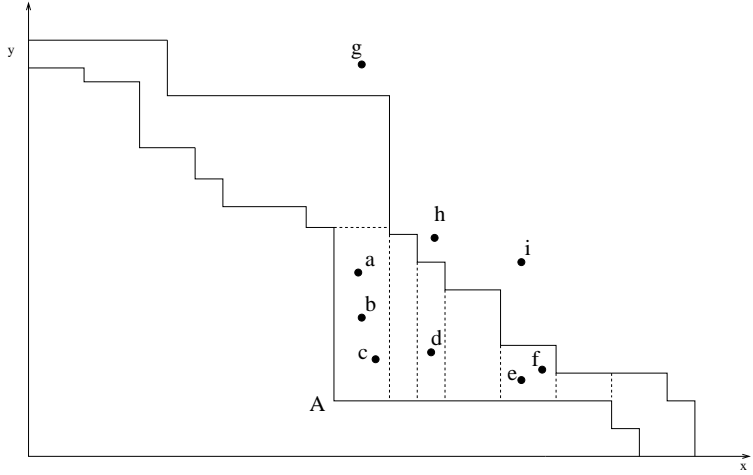

Figure 2: Construction of the dominance list for inward corner $A$. Points $a, b, c, d, e, f$ are "new" points. Point $g$ must be also included into the dominance list.

For each point $p$ that dominates $c_j$ and is stored in a dominance list of a $d$-successor of $c_j$ for $d \geq 4$, data structure $V^{\max}$ contains a point $(y_p, ind_l, z_p)$, such that $ind_l > \max_j$, $y_p \geq y_j$, and $z_p \geq z_j$. Again, for every $(y_p, ind_l, z_p)$ such that $ind_l > \max_j$, the $x$-coordinate $x_p$ of the corresponding point $p$ is greater than or equal to $x_j$. Hence, every element $(x_p, ind_l, z_p)$ of $V^{\max}$, such that $ind_l > \max_j$, $y_p \geq y_j$, and $z_p \geq z_j$ corresponds to a point that dominates $c_j$. All such elements can be found by a query to $V^{\max}$.

It remains to add the "new" points to the dominance list of $c_j$. Suppose that $c_j$ dominates inward corners $c'_1, c'_2, \ldots, c'_q$ on the previous ridge $R'_{i+1}$ and $c'_r = (x'_r, y'_r, z'_r)$ for $r = 1, \ldots, q$. We denote by $c_{j-1} = (x_{j-1}, y_{j-1}, z_j)$ and $c_{j+1} = (x_{j+1}, y_{j+1}, z_j)$ the 1-predecessor and the 1-successor of $c_j$. Let $x'_0 = x_j$, $y'_0 = y_{j-1}$, $x'_{q+1} = x_{j+1}$, $y'_{q+1} = y_j$. All "new" points can be found with $q+1$ queries $Q_1, Q_2, \ldots, Q_{q+1}$, where $Q_i = [x'_{i-1}, x'_i] \times [y'_i, y'_{i-1}] \times [z_i, +\infty)$ (see Fig. 2). The total number of queries that must be answered to find all "new" points for all inward corners of $R'_i$ is $O(n_i + n_{i-1})$, where $n_i$ is the number of inward corners on a ridge $R'_i$. Since $n_i = O(B^{1/3})$, we can answer all queries for a ridge $R'_i$ in $O(B^{2/3})$ I/Os by Lemma 6

When dominance lists for all inward corners of a ridge $R'_i$ are constructed, we update "points" in data structures $V^{\max}$ and $V^{\min}$. The list $L$ contains points $p$ reported by queries to $V^{\max}$ and $V^{\min}$. That is, the list $L$ contains all points $p$, such that $p$ is stored in the dominance list of some inward corner $c_f$ on ridge $R'_j$, $j > i$ and in the dominance list of some inward corner $c_h$ on ridge $R'_i$. Let $\min(p)$ and $\max(p)$ denote the minimal and the maximal $i$, such that $p$ is stored in the dominance list of the corner with index $i$. Using $O(N/B)$ additional space, we can determine whether a point $p$ belongs to the list $L$ and maintain for every point $p$ in $L$ the values $\min(p)$ and $\max(p)$.

For each point $p = (x_p, y_p.z_p)$ stored in $L$ we proceed as follows: If $\min(p) < ind_1$ for the corresponding point $(x_p, ind_1, z_p)$ stored in $V^{\max}$, we delete $(x_p, ind_1, z_p)$ from $V^{\max}$ and insert $(x_p, \min(p), z_p)$ into $V^{\max}$; if $\max(p) > ind_2$ for the corresponding point $(x_p, ind_2, z_p)$ stored in $V^{\min}$, we delete $(x_p, ind_2, z_p)$ from $V^{\min}$ and insert $(x_p, \max(p), z_p)$ into $V^{\max}$ For each "new" point $p$, we add $(x_p, \min(p), z_p)$ to $V^{\max}$ and $(x_p, \max(p), z_p)$ to $V^{\min}$. Since both $V^{\min}$ and $V^{\max}$ consist of just of a list of points (see Lemma 6) and the total number of points in both data structures is $O(B^{4/3})$, we can construct new versions of $V^{\min}$ and $V^{\max}$ in $O(B^{1/3})$ I/Os.

Since the total number of ridges is $O(B^{1/3})$, all dominance lists can be constructed in $O(B)$ I/Os.

## 6 Conclusion

In this paper we presented the first dynamic data structure that supports three-dimensional orthogonal range reporting queries in $O(\log_B^2 N + K/B)$ I/O operations. This query cost "matches" the query bound of the fastest internal memory data structure. The space usage of our data structure is quite comparable with the most space-efficient static data structure [2]. This is an interesting open question, whether the $O(\log_2^3 N)$ update cost can be significantly improved.

Using our approach, we can also obtain data structures that support special cases of range reporting queries; these data structures answer queries in $O(\log_B^2 N)$ I/Os, but use less space and support faster update operations than the data structure of Theorem 1. In particular, we can obtain:
(i) The data structure for $(1,1,1)$-sided queries (three-dimensional dominance queries) that uses $O((N/B)\log_2 N)$ blocks of space and supports updates in $O(\log_B^2 N)$ I/Os.
(ii) The data structure for $(1,1,2)$-sided queries that uses $O((N/B)\log_2 N \log_2 B)$ blocks of space and supports updates in $O(\log_2 N \log_B N)$ I/Os.
(iii) The data structure for $(2,1,2)$-sided queries that uses $O((N/B)\log_2 N \log_2^2 B)$ blocks of space and supports updates in $O(\log_2^2 N)$ I/Os.
The data structure (iii) is the result of Lemma 4. We obtain the results (i) and (ii) by replacing the data structures $D_v$, $D'_v$, and $E_v$ in the proof of Lemma 3 with data structures that support $(1,1,1)$-sided queries (resp. $(1,1,2)$-sided queries) on a set with $O(B^{4/3})$ points. Details will be given in the full version of this paper.